\documentclass[aps,apl,twocolumn,nowpacs,superscriptaddress,groupedaddress,amsmath,amssymb]{revtex4-1}
\usepackage {graphicx,epsfig,graphics,color}

\begin{document}


\title{Direct observation of magnetic phase coexistence and  magnetization reversal in a Gd$_{0.67}$Ca$_{0.33}$MnO$_{3}$ thin film}

\author{Jeehoon Kim}
\affiliation{Los Alamos National Laboratory, Los Alamos, NM 87545}
\email[Electronic mail: ]{jeehoon@lanl.gov}
\author{Nestor Haberkorn}
\affiliation{Los Alamos National Laboratory, Los Alamos, NM 87545}
\author{Leonardo Civale}
\affiliation{Los Alamos National Laboratory, Los Alamos, NM 87545}
\author{Evgeny Nazaretski}
\affiliation{Brookhaven National Laboratory, Upton, NY 11973}
\author{Paul Dowden}
\affiliation{Los Alamos National Laboratory, Los Alamos, NM 87545}
\author{Avadh Saxena}
\affiliation{Los Alamos National Laboratory, Los Alamos, NM 87545}
\author{J. D. Thompson}
\affiliation{Los Alamos National Laboratory, Los Alamos, NM 87545}
\author{Roman Movshovich}
\affiliation{Los Alamos National Laboratory, Los Alamos, NM 87545}


\begin{abstract}

We have investigated the ferrimagnetic domain structure in a
Gd$_{0.67}$Ca$_{0.33}$MnO$_{3}$ thin film using magnetic force
microscopy.  We observe clear signs of phase separation, with
magnetic islands embedded in a non-magnetic matrix. We also
directly visualize the reversal of magnetization of ferrimagnetic
domains as a function of temperature and attribute it to a change
in the balance of magnetization of anti-aligned Mn and Gd
sublattices.

\end{abstract}

\maketitle

Mixed-valent perovskite manganites A$_{1-x}$B$_{x}$MnO$_{3}$ (A
and B are rare-earth and divalent alkaline elements,
respectively), such as La-based manganites, have been studied
extensively in recent years.\cite {Jin 1994,Lee 2005,Wu 2006,Zhou
1999} These materials exhibit a colossal magnetoresistance (CMR)
effect for a wide range of dopings centered at x = 1/3 where the
double exchange mechanism is maximized.\cite{Lev 2004} The
resulting combination of fascinating physical phenomena and a
potential for technological applications has been the driving
force in sustaining high interest in these compounds.\cite {Jin
1994,Lee 2005,Wu 2006,Zhou 1999} Electronic inhomogeneity and
phase separation are ubiquitous in doped manganites, and the
resulting CMR effect is driven by percolative transport.\cite
{Dagotto 2004} CMR manifests itself by a dramatic drop in
resistivity and a discontinuous decrease in the equilibrium Mn-O
bond length at a first order phase transition in an applied
magnetic field.\cite {Zhang 2002, Goodenough 1997} Their complex
electronic structure and a variety of competing interactions lead
to a rich ensemble of ground states in this family of compounds.

In this Letter we report a low temperature magnetic force
microscopy (MFM) investigation of Gd$_{0.67}$Ca$_{0.33}$MnO$_{3}$
(GCMO), a compound with an insulating ferrimagnetic (FIM) ground
state. Compared to other ferromagnetic (FM) perovskite
manganites, GCMO exhibits a relatively low Curie temperature
({\it T$_C$}), and its small structural tolerance factor {\it t}
$<$ 0.89\cite {Hwang 1995,Jeffrey 1995} leads to a robust
insulating ground state. Magnetic properties of the system
reflect those of the two sublattices of Mn and Gd ions (see
below). The different temperature dependence of magnetization of
each of the two sublattices results in a change of sign of the
total magnetization as a function of temperature at a
characteristic compensation temperature {\it T$_{comp}$}, where
the Mn and Gd sublattices have magnetic moments of the same
magnitude and opposite direction.\cite {Hwang 1995,Jeffrey
1995,Pena 2002} A small tolerance factor, a structural distortion,
and the antiferromagnetic interaction between Gd and Mn
sublattices yield remarkable properties, such as a giant
magnetostrictive effect in a wide temperature range\cite {Correa
2011} and inhomogeneous FIM-like behavior with an exchange bias
effect close to {\it T$_{comp}$}.\cite {Nestor 2009} Low values
of the saturation magnetization ({\it M$_{S}$}) suggest phase
coexistence.\cite {Correa 2011, Nestor 2009} MFM studies
described below, with the wide range of field and temperature
employed, allow us to visualize the magnetic structure of GCMO
and provide direct evidence of phase separation. The
magnetization reversal at $T_{comp}$ of {\it each individual}
domain provides strong support for the scenario of anti-aligned
Mn and Gd sublattices with the Gd (Mn) magnetization dominating
below (above) $T_{comp}$.

The Gd$_{0.67}$Ca$_{0.33}$MnO$_{3}$ thin film was grown by
pulsed-laser deposition (PLD) on a SrTiO$_{3}$ (100) substrate
from a commercial target with the same chemical composition. The
substrate temperature was kept at 790 $^{\circ}$C in an oxygen
atmosphere at a pressure of 200 mTorr. After deposition, the
O$_{2}$ pressure was increased to 200 Torr, and the temperature
was decreased to room temperature at a rate of 30 $^{\circ}$C/min.
Bulk GCMO is an orthorhombic perovskite (Pbnm (no. 62); {\it a} =
5.52 \AA, {\it b} = 5.34 \AA, {\it c} = 7.50 \AA).\cite {Nestor
2009,Ma 2005} The GCMO film was examined by x-ray diffractometry,
and was found to be single phase with a (00{\it l}) orientation.
The lattice parameters ({\it a} = 5.55(2) \AA, {\it b} = 5.36(2)
\AA, and {\it c} = 7.50(1) \AA) were determined using (00{\it
l}), (200), and (020) reflections from a four-circle
diffractometer/goniometer. No additional peaks due to secondary
phases or different crystalline orientations were observed. The
rocking curve width around the (004) peak of the film was $\sim$
0.27$^{\circ}$. The film thickness of 45 nm was determined by a
low-angle x-ray reflectivity measurement with an angular
resolution of 0.005$^{\circ}$.

A Quantum Design SQUID magnetometer was used for measurements of
the global magnetization with the magnetic field oriented
perpendicular to the film surface. All localized MFM measurements
described in this Letter were performed in a home-built
low-temperature MFM apparatus.\cite {Nazaretski RSI 2009} MFM
images were taken in a frequency-modulated mode, with the
tip-lift height of 100 nm above the sample surface.  High
resolution SSS-QMFMR cantilevers,\cite {Nanosensors} magnetized
along the tip axis in a field of 3 T, were used for MFM
measurements; the external magnetic field was always applied
perpendicular to the film surface and parallel to the MFM tip.

\begin{figure}
\centering
\includegraphics [angle=0,width=7.5cm] {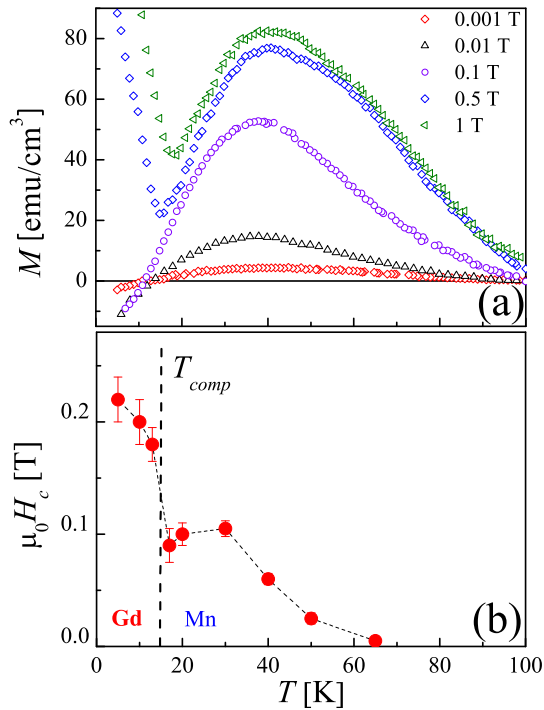}
\caption {\label{f:MT} (Color online) (a) Field-cooled {\it
M}({\it T}) curves in different magnetic fields ({\it H}). (b)
Coercive field ({\it H$_{c}$}) as a function of temperature
obtained from magnetic hysteresis loops.}
\end{figure}

Fig.~\ref{f:MT}(a) shows the field-cooled (FC) magnetization {\it
M} as a function of temperature at different values of applied
magnetic field {\it H}. The temperature dependence of
magnetization was discussed previously by Snyder {\it et al}.\cite
{Jeffrey 1995} GCMO undergoes a phase transition from
paramagnetic insulating to ferromagnetic insulating states,
associated with the ferromagnetic ordering of Mn cations, at {\it
T$_C$} $\approx$ 80 K. The local field due to FM order in the Mn
sublattice and the negative {\it f-d} exchange interaction on the
Gd spins force the moments on the Gd sublattice to be
anti-aligned to those in the Mn sublattice. The Mn sublattice
dominates the magnetization at high temperature, but the absolute
magnitude of magnetization of the Gd sublattice grows faster when
the temperature is reduced. Consequently, the total magnetization
{\it M} reaches a maximum value close to 50 K (see
Fig.~\ref{f:MT}), starts to decrease with decreasing temperature,
and goes toward zero at {\it T$_{comp}$} $\approx$ 15 K in low
fields ({\it T$_{comp}$} depends on {\it H}), where
magnetizations of the Mn and Gd sublattices compensate each
other. Below {\it T$_{comp}$} the local magnetization of Gd is
larger than that of Mn, $\mid M_{Gd}\mid > \mid M_{Mn}\mid$, and
the sign of the total magnetization is determined by the
direction of magnetization of the Gd sublattice. When the applied
magnetic field $H$ is below the coercive field $H_{c}$ of the
system at $T_{comp}$, the magnetization of the Gd sublattice is
locked in a direction opposite to the applied field, and the total
magnetization is negative below {\it T$_{comp}$}. For $H > H_{c}$
the magnetization is reversed immediately below $T_{comp}$,
producing a characteristic sharp kink and a V-shape in the data.
This sharp reversal of the change in magnetization (from
decreasing to increasing with decreasing temperature) is
facilitated by a strong decrease of $H_{c}$ at $T_{comp}$, as
shown in Fig.~\ref{f:MT}(b), which is determined on the basis of
an analysis of full hysteresis curves at different temperatures
(data not shown). The positive offset of $M$ at the kink at 0.5 T
and 1 T in Fig.~\ref{f:MT}(a) is due to a paramagnetic
background. All magnetic transition temperatures observed in the
film are in good agreement with the values previously reported
for bulk polycrystal and single crystal samples.\cite {Jeffrey
1995,Nestor 2009,Ma 2005} The data at 0.1 T has a clear kink as
it crosses $M = 0$ and $T_{comp}$, indicating that some small
number of the magnetic domains flip their orientation at
$T_{comp}$. This is consistent with the bulk measurements of $H_c
\approx  0.1$ T at $T_{comp}$, and points to coercive field in
this system being a local property, probably dependent on the
magnetic domain's size, shape, and environment.

\begin{figure}
\centering
\includegraphics [angle=-90,width=9.0cm] {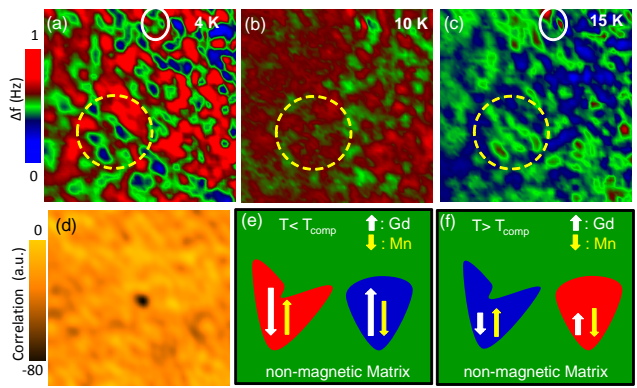}
\caption {\label{f:MFM-Temp}(Color online) (a)-(c) MFM images
acquired at different temperatures. Solid and dashed circles
represent the same sample area. (d) Cross-correlation map between
images shown in panels (a) and (c). The large negative value at
the center of the map signifies the anticorrelation between
images. (e) and (f) Schematical illustration of the temperature
evolution of phase-separated magnetic regions above and below
{\it T$_{comp}$} $\approx$ 12 K in 1 mT. The field of view in the
images ((a)-(d)) is 6 $\mu$m $\times$ 6 $\mu$m. Features on the
left side are broader than those on the right side because the
scan plane is not perfectly parallel to the sample surface.}
\end{figure}

\begin{figure}
\centering
\includegraphics [angle=-90,width=9.0cm] {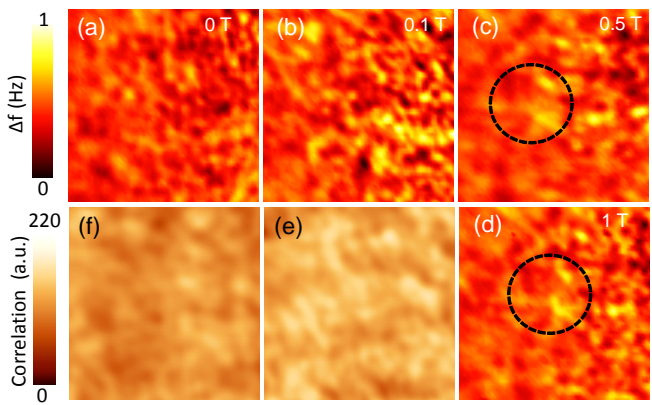}
\caption{\label{f:FC} (Color online) (a)-(d) Field-cooled MFM
images taken at 4 K in different magnetic fields. (e) and (f)
Cross-correlation maps between (a)-(b), and (b)-(c),
respectively; no correlation is observed. The field of view in
the images ((a)-(f)) is 6 $\mu$m $\times$ 6 $\mu$m. Dashed
circles correspond to the same sample area.}
\end{figure}

The MFM images depicted in Figs.~\ref{f:MFM-Temp}(a)-(c) were
taken sequentially at 4 K ({\it T}$<${\it T$_{comp}$}), 10 K
({\it T}$\approx${\it T$_{comp}$}), and 15 K ({\it T}$>${\it
T$_{comp}$}), respectively, in a magnetic field of 1 mT (below
$H_{c}$) applied above {\it T$_{C}$} (field-cooled data).  The
dashed circles in Figs.~\ref{f:MFM-Temp}(a)-(c) show the same
sample region (thermal drifts are negligible for images taken
below 15 K, see below). Regions of a non-zero magnetic signal,
either blue or red, change color as the temperature changes from
4 K to 15 K, but the green areas remain green in all images
(a)-(c) in Fig.~\ref{f:MFM-Temp}. The sample, therefore, is phase
separated into FIM (blue and red) and non-magnetic (green)
regions.\cite {Sun 2008} At 4 K (Fig.~\ref{f:MFM-Temp}(a)) Gd
dominates the magnetization of FIM domains, as depicted
schematically in Fig.~\ref{f:MFM-Temp}(e). Red islands in
Fig.~\ref{f:MFM-Temp}(a), therefore, represent those parts of the
sample where Gd magnetic moments point ``down", and the blue
regions are those with Gd magnetic moments pointing ``up". At 15 K
(Fig.~\ref{f:MFM-Temp}(c)) all of the red regions switch to blue,
signaling a reversal in their magnetization, as Mn magnetization
is dominant above $T_{comp}\approx 12$ K. This situation is
depicted schematically in Fig.~\ref{f:MFM-Temp}(f). The small
magnetic contrast across the sample at 10 K (see
Fig.~\ref{f:MFM-Temp}(b)) indicates almost equal magnetic
contributions of the anti-aligned Gd and Mn sublattices in FIM
regions near {\it T$_{comp}$}. In addition,
Fig.~\ref{f:MFM-Temp}(b) demonstrates the domains' breakup and a
reduction in their size close to {\it T$_{comp}$} (at 10 K). This
phenomenon is consistent with the exchange bias effect previously
observed in single crystals.\cite {Nestor 2009} The reduction of
the size of FIM domains close to {\it T$_{comp}$} also leads to a
decrease of the coercive field (see Fig.~\ref{f:MT}(b)).\cite {Ma
2005, Cullity}

Fig.~\ref{f:MFM-Temp}(d) shows a cross-correlation map between
images (a) and (c) and allows us to investigate qualitatively the
temperature evolution of magnetic domains in the sample. The
large negative value in the center of the cross-correlation map
demonstrates the anti-correlation between 4 K and 15 K
magnetization data in Fig.~\ref{f:MT}(a), indicating that red and
blue islands reverse their magnetization (and colors) upon the
temperature change from 4 K to 15 K. The central location of the
cross-correlation minimum also demonstrates the small thermal
drift in our MFM apparatus.

Figs.~\ref{f:FC}(a)-(d) show MFM images obtained at 4 K after
field-cooling the GCMO sample through $T_C$ in 0 T, 0.1 T, 0.5 T,
and 1 T applied fields. In order to understand the thermal and
field evolution of the sample's magnetization we evaluated
cross-correlation maps for these images. No correlation was
observed between data sets obtained at 0 T and 0.1 T (panels (a)
and (b)), as shown in Fig.~\ref{f:FC}(e), and 0.1 T and 0.5 T
(panels (b) and (c)), as shown in Fig.~\ref{f:FC}(f). The lack of
cross-correlation indicates significant evolution of the spin
magnetization due to the reversal process inside FIM clusters in
a field up to 0.5 T. On the other hand, magnetic domains imaged
in 0.5 T and 1 T FC experiments show a similar pattern,
suggesting saturation of the magnetization reversal process as
well as a clear phase separation between ferrimagnetic clusters
and the paramagnetic matrix. (Data taken at 3 T, not shown, are
similar to those at 1 T.) The lack of correlation between the
images (a)-(c) cannot be the result of thermal drift of the tip
position over the sample, as this was observed repeatedly to be
under 1 $\mu$m for our system (e.g., see panels (c) and (d)). The
magnetic-nonmagnetic phase coexistence could be attributed to
localized disorder or a localized strain distribution, similar to
observations in Y- and Pr-based manganites with a comparably low
tolerance factor [Y$_{2/3}$Ca$_{1/3}$MnO$_{3}$ ({\it t} $\sim$
0.884) and Pr$_{2/3}$Ca$_{1/3}$MnO$_{3}$ ({\it t} $\sim$
0.91)].\cite {Mathieu 2001, Radaelli 2001, Smol 2002, Saurel
2006} Results of x-ray diffraction measurements on our thin-film
sample, however, are close to those  on bulk samples and tend to
rule out a strain mechanism of phase separation.

In conclusion, we have performed MFM experiments on a
ferrimagnetic GCMO thin film and directly observed phase
separation in the sample, with magnetic (FIM) regions of
characteristic dimensions between 0.1 to 0.5 $\mu$m embedded in a
non-magnetic matrix. The behavior of magnetic regions is
consistent with the presence of anti-aligned Mn and Gd magnetic
sublattices, forming a FIM state. The observed magnetization
reversal in the FIM domains as a function of temperature, for
small external magnetic field, is consistent with the Mn
sublattice being dominant at $T > T_{comp} \approx 15$ K, but the
Gd sublattice (with magnetization locked to be antiparallel to a
small applied field) is dominant for $T <T_{comp}$. We attribute
the phase separation to localized disorder rather than a strained
state of the sample. These results will have significant bearing
on the potential utilization of GCMO and other related compounds
in magnetic memory device applications.

Work at LANL (sample fabrication, SQUID measurements, MFM, data
analysis, and manuscript preparation) was supported by the US
Department of Energy, Office of Basic Energy Sciences, Division
of Materials Sciences and Engineering. Work at Brookhaven (data
analysis and manuscript preparation) was supported by the US
Department of Energy under Contract No. DE-AC02-98CH10886. N.H.
is a member of CONICET (Argentina).

\end{document}